\documentstyle[12pt]{article}
\begin{document}
\def\theequation{\thesection.\arabic{equation}}
\newenvironment{proof}{\noindent {\em Proof. }}{\hfill$\Box$.\\}
\newtheorem{theo}{Theorem}[section]
\newtheorem{prop}[theo]{Proposition}
\newtheorem{lemm}[theo]{Lemma}
\newtheorem{coro}[theo]{Corollary}
\newtheorem{defi}[theo]{Definition}
\newtheorem{rema}[theo]{Remark}
\newcommand{\sectio}[1]{\section{#1}\setcounter{equation}{0}}
\newcommand{\e}{\label}
\newcommand{\r}[1]{(\ref{#1})}
\newcommand{\re}{\ref}
\newcommand{\k}{\ldots}
\newcommand{\api}{{\cal A}}
\newcommand{\bpi}{{\cal B}}
\newcommand{\cpi}{{\cal C}}
\newcommand{\lpi}{{\cal L}}
\newcommand{\ppi}{{\cal P}}
\newcommand{\spi}{{\cal I}}
\newcommand{\nb}{\mbox{{\bf N}}{}}
\newcommand{\rb}{\mbox{{\bf R}}{}}
\newcommand{\cb}{\mbox{{\bf C}}{}}
\newcommand{\jb}{\mbox{{\bf 1}}{}}
\newcommand{\cbg}{\cb_{*}}
\newcommand{\rbg}{\rb_{*}}
\newcommand{\al}{\alpha}
\newcommand{\ga}{\gamma}
\newcommand{\del}{\delta}
\newcommand{\eps}{\epsilon}
\newcommand{\es}{s}
\newcommand{\lam}{\lambda}
\newcommand{\si}{\sigma}
\newcommand{\de}{\Delta}
\newcommand{\la}{\Lambda}
\newcommand{\wb}{\bar w}
\newcommand{\g}{\Gamma}
\newcommand{\gd}{\dot\Gamma}
\newcommand{\poi}{Poincar\'e\ }
\newcommand{\qpg}{quantum \poi  group}
\newcommand{\qig}{quantum inhomogeneous group}
\newcommand{\be}{\begin{equation}}
\newcommand{\ee}{\end{equation}}
\newcommand{\bt}{\begin{theo}}
\newcommand{\et}{\end{theo}}
\newcommand{\bp}{\begin{prop}}
\newcommand{\ep}{\end{prop}}
\newcommand{\bl}{\begin{lemm}}
\newcommand{\el}{\end{lemm}}
\newcommand{\bc}{\begin{coro}}
\newcommand{\ec}{\end{coro}}
\newcommand{\bde}{\begin{defi}}
\newcommand{\ede}{\end{defi}}
\newcommand{\br}{\begin{rema}}
\newcommand{\er}{\end{rema}}
\newcommand{\bd}{\begin{proof}}
\newcommand{\ed}{\end{proof}}
\newcommand{\ba}{\begin{array}}
\newcommand{\ea}{\end{array}}
\newcommand{\btr}{\begin{trivlist}}
\newcommand{\etr}{\end{trivlist}}
\newcommand{\lra}{\longrightarrow}
\newcommand{\ot}{\otimes}
\newcommand{\op}{\oplus}
\newcommand{\tp}{\mbox{\raisebox{0.55mm}{\small{$\bigcirc$}}%
                 \hspace{-1.9ex}\small{$\top$}}}
\newcommand{\ti}{\mbox{\raisebox{0.55mm}{\small{$\bigcirc$}}%
                 \hspace{-1.9ex}\raisebox{0.65mm}{\small{$\bot$}}}}
\newcommand{\tu}{\mbox{\hspace{-0.4ex}%
                 \raisebox{0.55mm}{\scriptsize{$\bigcirc$}}%
                 \hspace{-1.45ex}\scriptsize{$\top$}\hspace{0.2ex}}}
\newcommand{\sd}{\rhd\mbox{\hspace{-2ex}}<}
\newcommand{\spa}{\mbox{{\rm{span}}}}
\newcommand{\po}{\mbox{{\rm{Poly}}}}
\newcommand{\id}{\mbox{{\rm{id}}}}
\newcommand{\mor}{\mbox{{\rm{Mor}}}}
\newcommand{\te}{\tilde}
\newcommand{\tb}{\te{\cal B}}
\newcommand{\iffi}{\Leftrightarrow}
\newcommand{\mod}{\mbox{{\rm{mod}}}}
\newcommand{\im}{\mbox{{\rm{im}}}}
\newcommand{\Sp}{\mbox{{\rm{Sp}}}}
\newcommand{\tr}{\mbox{{\rm{tr}}}}
\newcommand{\Irr}{\mbox{{\rm{Irr}}}}
\newcommand{\Rep}{\mbox{{\rm{Rep}}}}
\newcommand{\ov}{\overline}
\newcommand{\ite}{\item[]}
\newcommand{\hs}{\hspace{0.2ex}}
\newcommand{\pri}{{}^{\prime}}
\newcommand{\itemite}{\item[]\hspace{3ex}}
\newcommand{\Ren}{\mbox{{\rm{Re}}}}
\newcommand{\Imn}{\mbox{{\rm{Im}}}}
\title{On the classification of quantum Poincar\'e groups}
\author{P. Podle\'s${}^1$\thanks{On leave from
Department of Mathematical Methods in Physics,
Faculty of Physics, University of Warsaw,
Ho\.za 74, 00--682 Warszawa, Poland}
{}\thanks{This research was supported in part
by NSF grant DMS92--43893 and in part by Polish KBN grant No 2 P301 020 07}
and S. L. Woronowicz${}^2$\thanks{This research was supported by Polish KBN
grant No 2 P301 020 07}\\
${}^1$ Department of Mathematics, University of California\\
Berkeley CA 94720, USA\\
${}^2$ Department of Mathematical Methods in Physics\\
Faculty of Physics, University of Warsaw\\
Ho\.za 74, 00--682 Warszawa, Poland}
\date{March 1995\\{}}
\maketitle
\begin{abstract}
Using the general theory of \cite{INH},
\qpg s (without dilatations)
are described and investigated. The description contains a
set of numerical parameters which satisfy certain polynomial
equations. For most cases we solve them and give the classification of
\qpg s.
Each of them corresponds to exactly one quantum Minkowski space. The
Poincar\'e series of these objects are the same as in the classical case.
We also
classify possible $R$-matrices for the fundamental representation of the
group.
\end{abstract}
\setcounter{section}{-1}
\sectio{Introduction}

The Minkowski space  with the \poi group  acting on it is the area of the
quantum field theory. However, it is not known yet what is the area of a more
deep theory, which would involve also the gravitational effects. It was
suggested by many authors that it would be a quantum space. It means that
instead of functions on spacetime we would have elements of some
noncommutative algebra, called ``the algebra of functions on the quantum
space". On the other hand, such a quantum space should be in some sense
similar to the ordinary Minkowski space. The simplest models of such a
situation can be obtained by
choosing some properties of Minkowski space endowed
with the action of \poi group and classifying all quantum
groups and spaces which
satisfy
those properties.
There are many examples of quantum Poincar\'e groups, the corresponding
Minkowski spaces and R-matrices (cf e.g. \cite{L},
\cite{D}, \cite{S}, \cite{O},
\cite{M},
\cite{ChD}, \cite{GQ}
and remarks in \cite{INH} concerning these papers)
but such classification still doesn't exist.
Our aim is to provide it.
In Section~1 we define a quantum Poincar\'e group as a quantum group which is
built from any quantum Lorentz group \cite{WZ} and translations and satisfies
some natural properties. The corresponding commutation relations are
inhomogeneous and contain a set of parameters $H_{ABCD}$, $T_{ABCD}$.
Our scheme contains the examples provided in \cite{L}, \cite{ChD}, but
doesn't contain the examples of \cite{D}, \cite{S}, \cite{M}
(see however Remark {S3.9} of \cite{INH}) because we consider
quantum Poincar\'e groups without dilatations. Also the example \cite{O}
(formulated in the language of universal enveloping algebras)
has no corresponding object in our scheme (for $q\neq\pm1$).

It turns out
 that there are many quantum Lorentz groups which can be used in our
construction. However all of them correspond to $q=\pm1$. For each such quantum
Lorentz group (except the classical one and one more for $q=-1$
which are considered in Remark \re{r1.4}) we classify
all quantum Poincar\'e groups. We also provide the corresponding quantum
Minkowski spaces and R-matrices for the fundamental representation of the
quantum Poincar\'e group (for one family of considered \qpg s there is no
nontrivial R-matrix). The Poincar\'e series of the corresponding objects are
the same as in the classical case. The proofs of our results (using \cite{INH})
are contained in Section~2. In particular, the question of finding all \qpg s
is reduced to a set of polynomial equations for $H_{ABCD}$, $T_{ABCD}$
which we solve (in the indicated cases) using the computer MATHEMATICA
program.
Some results of the present paper were presented in
\cite{PW}. In \cite{Z} a similar classification is provided in the case of
Poisson manifolds and Poisson--Lie groups.

We use the terminology and results of \cite{INH}. Letter S means that we
make a reference to
\cite{INH}, e.g.  Theorem {S3.1} denotes Theorem 3.1
of \cite{INH}, (S1.2) denotes  equation (1.2) of \cite{INH}.
The small Latin indices $a,b,c,d,\ldots,$
belong to $\spi=\{0,1,2,3\}$ and
the capital Latin indices $A,B,C,D,\ldots,$ belong to $\{1,2\}$. We sum over
repeated indices which are not taken in brackets (Einstein's convention). The
number of elements in a set $B$ is $\#B$
or $|B|$. Unit matrix with dimension $N$ is
denoted by $\jb_N$, $\jb=\jb_2$. The Pauli matrices are given by
\[ \si_0=\jb_2,\quad
\si_1=\left(\ba{cc} 0&1 \\ 1&0 \ea\right),\quad
\si_2=\left(\ba{cc} 0&-i \\ i&0 \ea\right),\quad
\si_3=\left(\ba{cc} 1&0 \\ 0&-1 \ea\right). \]
If $V, W$ are vector spaces then $\tau_{VW}:V\ot W\lra W\ot V$ is given by
$\tau_{VW}(x\ot y) = y\ot x$, $x\in V$, $y\in W$. We often write $\tau$
instead of $\tau_{VW}$. We denote $\cbg=\cb\setminus\{0\}$,
$\rbg=\rb\setminus\{0\}$.

\sectio{Quantum \poi groups}

In this Section we define and (in almost all cases) classify quantum \poi
groups as objects having the properties of usual (spinorial) \poi group. The
proofs of the results are shifted to Section 2.

The (connected component of) vectorial \poi group
\[ \te P=SO_0(1,3)\sd\rb^4=\{(M,a): M\in SO_0(1,3), a\in\rb^4\} \]
has the multiplication $(M,a)\cdot(M',a')=(MM',a+Ma')$. By the \poi group we
mean spinorial \poi group (which is more important in quantum field theory
then $\te P$)
\[ P=SL(2,\cb)\sd\rb^4=\{(g,a): g\in SL(2,\cb), a\in\rb^4\} \]
with multiplication $(g,a)\cdot(g',a')=(gg',a+\lam_g(a'))$ where the double
covering $SL(2,\cb)\ni g\lra \lam_g\in SO_0(1,3)$ is given by
$\lam_g(x)_i\si_i=g(x_j\si_j)g^+$, $g\in SL(2,\cb)$, $x\in\rb^4$. The group
homomorphism $\pi:P\ni(g,a)\lra(\lam_g,a)\in\te P$ is also a double
covering. In particular, $(-\jb_2,0)\in P$ can be treated as rotation
about $2\pi$ which is trivial in $\te P$ but nontrivial in $P$ (it changes
the sign of wave functions for fermions). Both $P$ and $\te P$ act on
Minkowski space $M=\rb^4$ as follows $(g,a)x=(\lam_g,a)x=\lam_gx+a$, $g\in
SL(2,\cb)$, $a,x\in\rb^4$, and give affine maps preserving the scalar
product in $M$ (in more abstract setting we would treat $M$ as an affine
space
without distinguished $0$). Let us consider continuous functions $w_{AB}$,
$p_i$ on $P$ defined by
\[ w_{AB}(g,a)=g_{AB},\qquad p_i(g,a)=a_i. \]

We introduce
Hopf ${}^*$-algebra $\po(P)=(\bpi,\de)$ of polynomials on the \poi
group $P$ as the
${}^*$-algebra $\bpi$ with identity $I$ generated by $w_{AB}$
and $p_i$, $A,B=1,2$, $i\in\spi$ (according to Introduction,
$\spi=\{0,1,2,3\}$ in this Section) endowed with the
comultiplication $\de$ given by $(\de f)(x,y)=f(x\cdot y)$, $f\in\bpi$,
$x,y\in P$ ($f^*(x)=\ov{f(x)}$). In particular,
\be \de w_{CD}=w_{CF}\ot w_{FD}, \quad \de p_i=p_i\ot I+\la_{ij}\ot p_j,
\e{1.1} \ee
$p_i^*=p_i$, where
\be \la=V^{-1}(w\tp\wb)V, \qquad V=\left(\ba{cccc}
1&0&0&1 \\ 0&1&-i&0 \\ 0&1&i&0 \\ 1&0&0&-1 \ea \right). \e{1.2} \ee
In order to prove \r{1.1} we notice that
\[ (\de w_{CD})((g,a),(g',a'))=w_{CD}(gg',a+\lam_g(a'))=(gg')_{CD}= \]
\[ g_{CF}g'_{FD}=w_{CF}(g,a)w_{FD}(g',a')=(w_{CF}\ot w_{FD})((g,a),(g',a')),
\]
\[(\de p_i)((g,a),(g',a'))=p_i(gg',a+\lam_g(a'))=a_i+\lam_g(a')_i= \]
\[ a_i + (\lam_g)_{ij}a'_j = p_i(g,a)+\la_{ij}(g,a)p_j(g',a')=(p_i\ot I+
\la_{ij}\ot p_j)((g,a),(g',a')), \]
where we used the formulae $(\si_i)_{CD}=V_{CD,i}$ and
\[ V_{CD,i}(\lam_g)_{ij} = (\lam_g)_{ij}(\si_i)_{CD}= (g\si_j g^+)_{CD}= \]
\[ g_{CE}(\si_j)_{EF}(g^+)_{FD}=w_{CE}(g,a)V_{EF,j}{w_{DF}}^*(g,a)=    \]
\[ (w_{CE}\wb_{DF}V_{EF,j})(g,a)=V_{CD,i}\la_{ij}(g,a). \]
Since $\tau_{CD,EF}=\del_{CF}\del_{DE}$, we get
\be \bar V=\tau V \e{1.2a} \ee
and $\bar\la=\la$. We put $p=(p_i)_{i\in\spi}$. One can treat $w_{CD}$ as
continuous functions on the Lorentz group $L=SL(2,\cb)$ ($w_{CD}(g)=g_{CD}$,
$g\in L$). We define
Hopf ${}^*$-algebra $\po(L)=(\api,\de)$ of polynomials on
$L$ as ${}^*$-algebra with $I$ generated by all $w_{CD}$ endowed with $\de$
obtained by restriction of $\de$ for $\bpi$ to $\api$. Clearly $w$ and $\la$
are representations of $L$. It is easy to check that
\begin{trivlist}
\item[1.]  $\bpi$ is generated as algebra by $\api$
and the elements $p_i$, $i\in\spi$.
\item[2.] $\api$ is a Hopf ${}^*$-subalgebra of $\bpi$.
\item[3.] $\ppi=\left(\ba{cc}\la&p\\0&I\ea\right)$ is a representation of
$\bpi$ where $\la$ is given by \r{1.2}.
\item[4.] There exists $i\in\spi$ such that $p_i\not\in\api$.
\item[5.] $\g\api\subset\g$ where $\g=\api X+\api$, $X=\spa\{p_i:i\in\spi\}$.
\item[6.] The left $\api$-module $\api\cdot\spa\{p_ip_j, p_i, I:
i,j\in\spi \}$ has a free basis consisting of $10+4+1$ elements.
\end{trivlist}
( 5. and 6. follow from the relations $p_ia=ap_i$, $p_ip_j=p_jp_i$,
$a\in\api$, and elementary computations, a free basis is given by $\{p_ip_j,
p_i, I: i\leq j, i,j\in\spi \}$). According to \cite{WZ}, $\po(L)$
satisfies:
\begin{trivlist}
\item[i.] $(\api,\de)$ is a Hopf ${}^*$-algebra such that $\api$ is generated
(as
${}^*$-algebra) by matrix elements of a two--dimensional representation $w$
\item[ii.] $w\tp w\simeq I\oplus w^1$ where $w^1$ is a representation
\item[iii.] the representation $w\tp\wb\simeq\wb\tp w$ is irreducible
\item[iv.] if $\api',\de',w'$ satisfy i.--iii. and there exists
Hopf ${}^*$-algebra
epimorphism $\rho:\api'\lra\api$ such that $\rho(w')=w$ then $\rho$
is
an isomorphism (the universality condition)
\end{trivlist}

We say \cite{WZ} that $H$ is a quantum Lorentz group if $\po(H)=(\api,\de)$
satisfies i.--iv..

\bde\e{d1.1} We say that $G$ is a \qpg\ if
Hopf ${}^*$-algebra $\po(G)=(\bpi,\de)$ satisfies the
conditions 1.--6. for some quantum Lorentz group $H$ with
$\po(H)=(\api,\de)$ and a representation $w$ of $H$. \ede

\br\e{r1.1} The condition 5. follows from $\ppi\tp w\simeq w\tp\ppi$,
$\ppi\tp\wb\simeq\wb\tp\ppi$, while 6. is suggested by the requirement
$W(\ppi\tp\ppi)=(\ppi\tp\ppi)W$ for a ``$\tau$-like'' matrix $W$ (cf Theorem
\re{t1.8}). Moreover, the condition 4. is superfluous (it follows from the
condition 6. and Proposition {S0.1}).\er

\br\e{r1.2} Different choices of $(H,w)$ can give $*$-isomorphic
$\bpi$.\er

\bt\e{t1.2} Let $G$ be a \qpg, $\po(G)=(\bpi,\de)$. Then $\api$ is
linearly
generated by matrix elements of irreducible representations of $G$,
so
$\api$ is uniquely determined. Moreover, we can choose $w$ in such a way that
$\api$ is the universal
${}^*$-algebra generated by $w_{AB}$, $A,B=1,2,$ satisfying
\be (w\tp w)E=E, \e{1.3}\ee
\be E'(w\tp w)=E', \e{1.4} \ee
\be X(w\tp\wb)=(\wb\tp w)X, \e{1.5} \ee
where $X=\tau Q'$ and
\begin{trivlist}
\item[1)] $E=e_1\ot e_2-e_2\ot e_1$, $E'=-e^1\ot e^2+e^2\ot e^1$,
\[ Q'=\left[ \ba{cccc} t^{-1} & 0 & 0 & 0 \\ 0 & t & 0 & 0\\ 0 & 0 & t & 0\\
             0 & 0 & 0 & t^{-1}\ea\right], \quad 0<t\leq1,\quad \mbox{or} \]
\item[2)] \[ E,E' \mbox{ as above },\quad
 Q'=\left[ \ba{cccc} 1 & 0 & 0 & 1 \\ 0 & 1 & 0 & 0\\ 0 & 0 & 1 & 0\\
             0 & 0 & 0 & 1\ea\right],\quad \mbox{or} \]
\item[3)] $E=e_1\ot e_2-e_2\ot e_1+e_1\ot e_1$, $E'=-e^1\ot e^2+e^2\ot
e^1+e^2\ot e^2$,
 \[ Q'=\left[ \ba{cccc} 1 & 0 & 0 & r \\ 0 & 1 & 0 & 0\\ 0 & 0 & 1 & 0\\
             0 & 0 & 0 & 1\ea\right],\quad r\geq0,\quad \mbox{ or } \]
\item[4)] \[ E,E' \mbox{ as above },\quad
 Q'=\left[ \ba{cccc} 1 & 1 & 1 & 0 \\ 0 & 1 & 0 & -1\\ 0 & 0 & 1 & -1\\
             0 & 0 & 0 & 1\ea\right],\quad  \mbox{ or } \]
\item[5)] $E=e_1\ot e_2+e_2\ot e_1$, $E'=e^1\ot e^2+e^2\ot e^1$,
 \[ Q'=i\left[ \ba{cccc} t^{-1} & 0 & 0 & 0 \\ 0 & -t & 0 & 0\\
 0 & 0 & -t & 0\\
             0 & 0 & 0 & t^{-1}\ea\right],\quad 0<t\leq1,\quad
	     \mbox{ or } \]
\item[6)] \[ E,E' \mbox{ as above },
 Q'=i\left[ \ba{cccc} 1 & 0 & 0 & 1 \\ 0 & -1 & 0 & 0\\ 0 & 0 & -1 & 0\\
             0 & 0 & 0 & 1\ea\right],\quad \mbox{ or } \]
\item[7)] \[ E,E' \mbox{ as above },\quad
 Q'=i\left[ \ba{cccc} r & 0 & 0 & s \\ 0 & -r & s & 0\\ 0 & s & -r & 0\\
             s & 0 & 0 & r\ea\right], \]
\[ r=(t+t^{-1})/2,\quad s=(t-t^{-1})/2,\quad 0<t<1, \]
\end{trivlist}
$e_1=\left(\ba{c}1\\0\ea\right)$, $e_2=\left(\ba{c}0\\1\ea\right)$,
$e^1=\left(\ba{cc}1&0\ea\right)$, $e^2=\left(\ba{cc}0&1\ea\right)$. Moreover,
all the above triples $(E,E',Q')$ give nonisomorphic $(\api,\de)$. We can
(and will)
choose $p_i$ in such a way that $p_i^*=p_i$.\et

In the following we assume that $G$ is a \qpg, $\po(G)=(\bpi,\de)$ and $w,p$
are as in Theorem \re{t1.2}. We set $q=q^{1/2}=1$ in the cases 1)--4),
$q=-1$,
$q^{1/2}=i$ in the cases 5)--7), $\es=\pm1$,
$L=\es q^{1/2}(\jb^{\ot 2}+q^{-1}EE')$, $\te L=q\tau L\tau$,
$G=(V^{-1}\ot\jb)(\jb\ot X)(L\ot\jb)(\jb\ot V)$,
$\te G=(V^{-1}\ot\jb)(\jb\ot\te L)(X^{-1}\ot\jb)(\jb\ot V)$,
$R=(V^{-1}\ot V^{-1})(\jb\ot X\ot\jb)(L\ot\te L)(\jb\ot X^{-1}\ot\jb)
(V\ot V)$.

\bt\e{t1.3} $\bpi$
is the universal ${}^*$-algebra with $I$ generated by $w_{AB}$ and $p_i$
satisfying \r{1.3}, \r{1.4}, \r{1.5} and
\be p_ia=(a*f_{ij})p_j+a*\eta_i-\la_{ij}(\eta_j*a),\quad a\in\api,\e{1.6}\ee
\be(R-\jb^{\ot 2})_{kl,ij}(p_ip_j-\eta_i(\la_{js})p_s+
T_{ij}-\la_{im}\la_{jn}T_{mn})=0,\e{1.7}\ee
\be p_i^*=p_i,\e{1.8}\ee
where $f=(f_{ij})_{i,j\in\spi}$, $\eta=(\eta_i)_{i\in\spi}$ and
$T=(T_{ij})_{i,j\in\spi}$ are uniquely determined by $\es=\pm1$,
$H_{EFCD},T_{EFCD}\in\cb$ and the following properties:
\begin{trivlist}
\item[a)] $\api\ni a\lra\rho(a)=\left(\ba{cc} f(a) & \eta(a) \\ 0 &
\eps(a)\ea\right)\in M_5(\cb)$ is a unital homomorphism
\item[b)] $\rho(a^*)=\ov{\rho(S(a))},\qquad a\in\api,$
\item[c)] $f_{ij}(w_{CD})=G_{iC,Dj}$,\quad
$\eta_i(w_{CD})=V^{-1}_{i,EF}H_{EFCD}$,\quad
$T_{ij}=(V^{-1}\ot V^{-1})_{ij,EFCD}T_{EFCD}$.
\end{trivlist}

The $*$-Hopf structure in $\po(G)$ is determined by:
\[ \de w=w\ti w,\quad  \de\wb=\wb\ti\wb,\quad  \de p=p\ti I+\la\ti p, \]
\[ \eps(w)=\jb,\quad \eps(\wb)=\jb,\quad, \eps(p)=0, \]
\[ S(w)=w^{-1},\quad  S(\wb)=\wb^{-1},\quad S(p)=-\la^{-1}p. \]
Quantum Poincar\'e groups corresponding to different $\es$ are
nonisomorphic.\et

\bt\e{t1.4} For each case in Theorem \re{t1.2} and each $\es$ (except
the case 1), $\es=1$, $t=1$ and the case 5), $\es=\pm1$, $t=1$) we list $H$
and
$T$ giving (via formulae in Theorem \re{t1.3}) all nonisomorphic \qpg s $G$:
\btr
\item[] 1), $\es=-1$, $t=1$:\\
\be \left.\ba{rcl} H_{EFCD} & = & 0,\\\\
T_{EFCD} & = & V_{EF,i}V_{CD,j}T_{ij},
\ea\right\}\e{1.8a} \ee
where
\itemite a) $T_{03}=-T_{30}=ia$, $T_{12}=-T_{21}=ib$, other $T_{ij}$ equal
$0$, $a=\cos\phi$, $b=\sin\phi$ (one parameter family for
$0\leq\phi\leq\pi/2$) or
\itemite b) $T_{02}=T_{12}=i$, $T_{20}=T_{21}=-i$, other $T_{ij}$ equal
$0$, or
\itemite c) all $T_{ij}$ equal $0$.
\item[] 1), $\es=\pm1$, $0<t<1$:
\be\left.\ba{c} T_{1122}=ia,\quad T_{1221}=b,\\\\
                T_{2112}=-b,\quad T_{2211}=-ia,\\\\
\mbox{ all } H_{EFCD} \mbox{ and other } T_{EFCD}
\mbox{ equal } 0 \mbox{ and }
\ea\right\}\e{1.9}\ee
\itemite a) $a=\cos\phi$, $b=\sin\phi$ (one parameter family for
$0\leq\phi<\pi$) or
\itemite b) $a=b=0$.
\item[] 2), $\es=1$:\\
\noindent the first case:
\be\left.\ba{c} H_{1111}=-(a+bi),\quad H_{1122}=a+bi,\quad H_{2112}=-2bi,\\\\
T_{2111}=c-di,\quad T_{1211}=-c-di,\\\\
 T_{1121}=-c+di,\quad
T_{1112}=c+di,\\\\
\mbox{ other } H_{EFCD} \mbox{ and } T_{EFCD} \mbox{ equal } 0 \mbox{ and }
\ea\right\}\e{1.10}\ee
\itemite a) $a=1$, $c=d=0$ (one parameter family for $b\in\rb$) or
\itemite b) $a=0$, $b=1$, $d=0$ (one parameter family for $c\geq0$);

\noindent the second case:
\be\left.\ba{c}
H_{1212}=a+bi,\quad T_{2112}=(a^2+b^2)/2,\quad T_{2111}=c-di,\\\\
T_{1221}=-(a^2+b^2)/2,\quad T_{1211}=-c-di,\quad T_{1121}=-c+di,\quad\\\\
T_{1112}=c+di,\quad T_{1111}=-(a^2+b^2)/2,\\\\
\mbox{ other } H_{EFCD} \mbox{ and } T_{EFCD} \mbox{ equal } 0 \mbox{ and }
\ea\right\}\e{1.11}\ee
\itemite a) $a=1$, $b=0$, $c=r\cos\phi$, $d=r\sin\phi$ (two parameter family
for $r>0$, $0\leq\phi<\pi/2$ or $r=\phi=0$) or
\itemite b) $a=b=0$, $c=1$, $d=0$, or
\itemite c) $a=b=c=d=0$.
\item[] 2), $\es=-1$, \r{1.10} and
\itemite a) $a=b=0$, $c=1$, $d=0$, or
\itemite b) $a=b=c=d=0$.
\item[] 3), $\es=\pm1$, $r\geq0$, all $H_{EFCD}$ and $T_{EFCD}$ equal $0$.
\item[] 4), $\es=1$,
\be\left.\ba{c}
H_{2212}=-2bi,\quad H_{2122}=-bi,\quad H_{2112}=a-bi,\\\\
H_{2111}=bi,\quad H_{1222}=bi,\quad H_{1212}=a,\quad H_{1211}=-bi,\\\\
H_{1121}=-2bi,\quad H_{1112}=3bi/4,\quad H_{1111}=-4bi,\\\\
T_{1112}=9b^2/8+3abi/2,\quad T_{1121}=-9b^2/8+3abi/2,\\\\
T_{1211}=-9b^2/8-3abi/2,\quad T_{1221}=3b^2/2,\\\\
T_{2111}=9b^2/8-3abi/2,\quad T_{2112}=-3b^2/2,\\\\
\mbox{ other } H_{EFCD} \mbox{ and } T_{EFCD} \mbox{ equal } 0 \mbox{ and }
\ea\right\}\e{1.12}\ee
\itemite a) $a=\cos\phi$, $b=\sin\phi$ (one parameter family for
$0\leq\phi<\pi$) or
\itemite b) $a=b=0$.
\item[] 4), $\es=-1$, all $H_{EFCD}$ and $T_{EFCD}$ equal $0$.
\item[] 5), $\es=\pm1$, $0<t<1$,
\be\left.\ba{c}
T_{1122}=ia,\quad T_{1221}=b,\quad T_{2112}=-b,\quad T_{2211}=-ia,\\\\
\mbox{ all } H_{EFCD} \mbox{ and other } T_{EFCD} \mbox{ equal } 0
\mbox{ and }\ea\right\}\e{1.13}\ee
\itemite a) $a=\cos\phi$, $b=\sin\phi$ (one parameter family for
$0\leq\phi<\pi$) or
\itemite b) $a=b=0$.
\item[] 6), $\es=1$,
all $H_{EFCD}$ and $T_{EFCD}$ equal 0.
\item[] 6), $\es=-1$:\\
the first case:
\be\left.\ba{c} H_{1111}=-(a+bi),\quad H_{1122}=a+bi,\quad H_{2112}=-2bi\\\\
\mbox{ other } H_{EFCD} \mbox{ and all } T_{EFCD} \mbox{ equal } 0
\mbox{ and }
\ea\right\}\e{1.14}\ee
\itemite a) $a=\cos\phi$, $b=\sin\phi$ (one parameter family for
$0\leq\phi<\pi$) or
\itemite b) $a=b=0$;\\
\noindent the second case:\\
\be\left.\ba{c} H_{1212}=a+bi,\quad T_{1111}=-\frac12(a^2+b^2),\\\\
T_{1221}=-(a^2+b^2)/2,\quad
T_{2112}=(a^2+b^2)/2,\\\\
\mbox{ other } H_{EFCD} \mbox{ and } T_{EFCD} \mbox{ equal } 0
\mbox{ and }
\ea\right\}\e{1.15}\ee
$a=1$, $b=0$.
\item[] 7), $\es=\pm1$, $0<t<1$, all $H_{EFCD}$ and $T_{EFCD}$ equal
$0$. \etr\et

\br The classical Poincar\'e group is obtained in the case 1), $s=1$, $t=1$,
$H=0$, $T=0$. The quantum Poincar\'e group of \cite{L} corresponds
(in spinorial setting) to
1), $s=1$, $t=1$,
\[ H_{1111}=-H_{1122}=\frac12H_{1221}=\frac12H_{2112}=-H_{2211}=H_{2222}=ih/2,
\ \ h\in\rb, \]
other $H_{EFCD}$ and all $T_{EFCD}$ equal $0$. The quantum Poincar\'e
group of \cite{ChD} corresponds to 1), $s=1$, $t>0$, $H=0$, $T=0$
($t$ is denoted by $q$ there). The so
called soft deformations correspond to 1), $s=\pm1$, $t=1$, $H=0$,
$T_{ab}=-T_{ba}\in i\rb$.\er

\br\e{r1.4} In the remaining cases 1), $s=1$, $t=1$ and 5), $s=\pm1$, $t=1$,
one can consider $T_{mn}$ defined as in Theorem \re{t1.3} and
\[ Z_{ij,k}=\eta_i(\la_{jk})=V^{-1}_{i,AB}V^{-1}_{j,CD}(H_{ABCE}\del_{DF}-
\overline{H_{BADF}}\del_{CE})V_{EF,k} \]
(then $H_{ABCE}=\frac12 V_{AB,i}V_{CD,j}Z_{ij,k}V^{-1}_{k,ED}$).
In the case 1), $s=1$, $t=1$ a pair $(Z,T)$ corresponds to a
\qpg\ if and only if
\be T_{mn}=-T_{nm}\in i\rb,\qquad Z_{ij,s}g_{sk}=-Z_{ik,s}g_{sj}
\in i\rb,\e{1.15'}\ee
\be\left.\ba{c}
\{[(\tau-\jb^{\ot 2})\ot\jb][(\jb\ot Z)Z-(Z\ot\jb)Z]\}_{ijm,n}=\\
-\frac14 t_0(\del_{in}g_{jm}-\del_{jn}g_{im}),\quad t_0\in\rb,\\
A_3(Z\ot\jb)T=0\ea\right\}\e{1.15''}\ee
where $g_{00}=1$, $g_{11}=g_{22}=g_{33}=-1$, other $g_{ij}=0$,
\[ A_3=\jb\ot\jb\ot\jb-\tau\ot\jb-\jb\ot\tau+(\tau\ot\jb)(\jb\ot\tau)+ \]
\[ (\jb\ot\tau)(\tau\ot\jb)-(\tau\ot\jb)(\jb\ot\tau)(\tau\ot\jb) \]
is the classical (not normalized) antisymmetrizer.
In the case 5), $s=\pm1$, $t=1$ in addition to these conditions we assume
\[ T_{i_1i_2}=0 \mbox{ for } \#\{k:\ i_k\in\{1,2\}\}=1, \]
\[ Z_{i_1i_2,i_3}=0 \mbox{ for } (-1)^{\#\{k:\ i_k\in\{1,2\}\}}=s \]
and get in that way all \qpg s (up to isomorphism but not necessarily
nonisomorphic).

Let us set $\eta=g$, $a=-iT_{mn}e_m\wedge e_n$,
$b=-iZ_{ij,s}g_{sk}e_i\wedge\Omega_{j,k}$ and $c=0$ (see \cite{Z}).
Then \r{1.15''} (using \r{1.15'}) is equivalent to (3)-(4) of \cite{Z}
where $t_0$ is identified with $t$ of (3)-(4) of \cite{Z}.
Thus the table in \cite{Z} gives many examples of \qpg s (cf also the
remarks at the end of \cite{Z}). The proofs of these statements involve
the above formulae and the results obtained in the proof of Theorem
\re{t1.4} (with $\lam=-\frac12 t_0$ in the case 1), $s=1$, $t=1$
and $\lam=-\frac12 it_0$ in the case 5), $s=\pm1$, $t=1$).\er

We denote by $d_n$ the number of monomials of $n$th degree in $4$ variables,
\[ d_n=\#\{(a,b,c,d)\in\nb^{\ot 4}:\ \ a+b+c+d=n\}. \]

\bt\e{t1.5} Let $\bpi$ correspond to a \qpg\ $G$ and $\api, w, p$ be as in
Theorem \re{t1.2}. We set
\[ \bpi^N=\api\cdot\spa\{p_{i_1}\cdot\k\cdot p_{i_n}: i_1,\k,i_n\in\spi,\quad
n=0,1,\k,N \}. \]
Then $\bpi^N$ is a free left $\api$-module and
$\dim_{\api}\bpi^N=\sum_{n=0}^N d_{n}.$ \et

We denote by $l:P\times M\lra M$ the action of \poi group on Minkowski space,
$\cpi=\po(M)$ denotes the unital algebra generated by coordinates $x_i$
($i\in\spi$) of the Minkowski space $M=\rb^4$. The only relations in $\cpi$
are
$x_ix_j=x_jx_i$. The coaction $\Psi:\cpi\lra\api\ot\cpi$ and $*$ in $\cpi$
are
given by $(\Psi f)(x,y)=f(l(x,y))$,
$f^*(y)=\ov{f(y)}$, $x\in P$, $y\in M$.

Let $x=(g,a)\in P$, $y\in M$, $f\in\cpi$. One has
\[ (\Psi x_i)((g,a),y)=x_i(\lam_g y+a)=(\lam_g y)_i+a_i=
(\lam_g)_{ij}y_j+a_i= \]
\[ \la_{ij}(g,a)x_j(y)+p_i(g,a)= (\la_{ij}\ot x_j+p_i\ot I)((g,a),y),
\mbox{ hence } \]
\be \Psi x_i=\la_{ij}\ot x_j+p_i\ot I.\e{1.16}\ee
One gets

\btr \item[6)] $\cpi$ is a unital
${}^*$-algebra generated by $x_i$, $i\in\spi$,
and $\Psi :\cpi\lra \bpi\ot\cpi$ is a unital ${}^*$-homomorphism such that
$(\eps\ot\id)\Psi=\id$, $(\id\ot\Psi)\Psi=(\de\ot\id)\Psi$,
$x_i^*=x_i$ and \r{1.16}
holds.\etr

Let $\Psi W\subset\api\ot W$ for a linear subspace $W\subset\cpi$, $f\in W$,
$y,a\in\rb^4$. Then
\[ f(y+a)=f(l((e,a),y))=(\Psi f)((e,a),y)=(\Psi f)((e,0),y)=
f(l((e,0),y))=f(y) \]
( $k(e,a)=k(e,0)$ for $k\in\api$), $f=f(0)I\in\cb I$
(in fact we have used the translation homogeneity of $M$). Therefore

\btr \item[7)] if $\Psi W\subset \api\ot W$
for a linear subspace $W\subset\cpi$ then
$W\subset\cb I$.\etr

Let us consider $(\cpi',\Psi')$ which also satisfies 6)--7) for some
${x_i}'\in\cpi'$. Then
\[
\Psi({x_i}'{x_l}'-{x_l}'{x_i}')=\la_{ij}\la_{lm}\ot({x_j}'{x_m}'-
{x_m}'{x_j}'). \]
Setting $W=\spa\{ {x_i}'{x_l}'-{x_l}'{x_i}': i,l\in\spi\}$ and using 7), one
gets ${x_i}'{x_l}'-{x_l}'{x_i}' = a_{il}I$, $a_{il}\in\cb$. Thus
$a=(a_{il})_{i,l\in\spi}$ is an invariant vector of $\la\tp\la$, i.e.
$a=c\cdot
g$ where $c\in\cb$, $g_{00}=1$, $g_{11}=g_{22}=g_{33}=-1$, $g_{ij}=0$ for
$i\neq j$. But $a_{il}=-a_{li}$,
hence $c=0$,
${x_i}'{x_l}'={x_l}'{x_i}'$ and we fix the proper choice of $(\cpi,\Psi)$
by means of
\btr \item[8)] if $(\cpi',\Psi')$ also satisfies 6)--7) for some
${x_i}'\in\cpi'$ then
there exists a unital ${}^*$-homomorphism $\rho:\cpi\lra\cpi'$ such that
$\rho(x_i)={x_i}'$ and $(\id\ot\rho)\Psi=\Psi'\rho$ (universality of
$(\cpi,\Psi)$).
\etr

\bde\e{d1.6} We say that $(\cpi,\Psi)$ describes a quantum Minkowski space
associated with a \qpg\ $G$, $\po(G)=(\bpi,\de)$, if 6)--8) are
satisfied.\ede

\br This definition doesn't depend on the choice of
$\la$ (see Proposition {S5.7}).\er

\bt\e{t1.7} Let $G$ be a \qpg\ with $w,p$ as in Theorem
\re{t1.2}.
Then there exists a unique (up to a ${}^*$-isomorphism) pair $(\cpi,\Psi)$
 describing
associated Minkowski space:

$\cpi$ is the
universal unital ${}^*$-algebra generated by $x_i$, $i=0,1,2,3$, satisfying
${x_i}^*=x_i$ and
\be (R-\jb^{\ot 2})_{ij,kl}(x_k x_l-\eta_k(\la_{lm})x_m+T_{kl})=0,\e{1.17}\ee
and $\Psi$ is given by $\r{1.16}$. Moreover,
\be \dim\cpi^N=\sum^N_{n=0} d_n, \e{1.18}\ee
where $\cpi^N=\spa\{x_{i_1}\cdot\k\cdot x_{i_n}:\ \ i_1,\k,i_n\in\spi,\ \
n=0,1,\k,N\}.$\et

We set $m=(V^{-1}\ot V^{-1})(\jb\ot X\ot\jb)(E\ot\tau E)$,
$Z_{ij,k}=\eta_i(\la_{jk})$,
\be R_P=\left(\ba{cccc} R & Z & -R\cdot Z & (R-\jb^{\ot 2})T \\
                        0 & 0 &      \jb  &     0      \\
			0 &\jb&      0    &     0      \\
			0 & 0 &      0    &     1      \ea\right),
m_P=\left(\ba{cccc}     0 & 0 & 0 & m\\
                        0 & 0 & 0 & 0\\
			0 & 0 & 0 & 0\\
			0 & 0 & 0 & 0\ea\right).\e{1.18m}\ee

\bt\e{t1.8} Let $G$ be a \qpg\ with $w,p$ as in Theorem \re{t1.2}. Then
\btr
\item[1)] $\mor(\ppi\tp\ppi,\ppi\tp\ppi)=\cb\id\op\cb R_P\op\cb m_P$.
\item[2)] Let us consider the cases listed in Theorem \re{t1.4}. Then
$W\in\mor(\ppi\tp\ppi,\ppi\tp\ppi)$ and
\be (W\ot\jb)(\jb\ot W)(W\ot\jb)=(\jb\ot W)(W\ot\jb)(\jb\ot W)\e{1.18a}\ee
if and only if
\itemite a) $W=x\cdot\id$ ($x\in\cbg$) or
\itemite b) $W=y\cdot R_P+z\cdot m_P$ ($y,z\in\cb$, for 4), $s=1$, $b\neq0$
one must have $y=0$).

Those  $W$ are invertible if and only if we have the case a) or b) with
$y\neq0$.\etr\et

\sectio{Proof of the classification}

In this Section we
prove the Theorems of Section~1.

Let $H$ be a quantum Lorentz group, i.e. $Poly(H)=(\api,\de)$ satisfies
the conditions i.-iv. of Section 1.
According to \cite{WZ}, we can choose $w$ in such a way that $\api$ is the
universal
${}^*$-algebra generated by $w_{AB}$, $A,B=1,2$, satisfying \r{1.3}--\r{1.5},
where $X=\tau Q'$, $Q'=\al Q$ and
\btr
\item[1)] $E=e_1\ot e_2-q e_2\ot e_1$, $E'=-q^{-1}e^1\ot e^2+e^2\ot e^1$, $Q$
is given by (13)--(19) of \cite{WZ}, $q\in\cb\backslash\{0,i,-i\}$, or
\item[2)] $E=e_1\ot e_2-e_2\ot e_1+e_1\ot e_1$,
$E'=-e^1\ot e^2+e^2\ot e^1+e^2\ot e^2$, $Q$ is given by (20)-(21) of
\cite{WZ}, we set $q=1$ in that case,
\etr
$e_1=\left(\ba{c}1\\0\ea\right)$, $e_2=\left(\ba{c}0\\1\ea\right)$,
$e^1=(1, 0)$, $e^2=(0, 1)$ (due to remarks before formula (1) in
\cite{WZ}, $E'E\neq0$ which means $q\neq\pm i$). In all these cases $X$ is
invertible, $\de$ is given by $\de w_{ij}=w_{ik}\ot w_{kj}$ and $(\api,\de)$
corresponds to a quantum Lorentz group. The numbers $\al\neq0$ are not
essential now and are chosen in such a way that
\be (X\ot\jb)(\jb\ot X)(E\ot\jb)=\jb\ot E \e{4.1}\ee
(see (5) of \cite{WZ}). Then (we use (6) of \cite{WZ} and direct
computations)
\be \tau\bar X\tau=\beta^{-1}X \e{4.2}\ee
for some $\beta\in\{1,-1,i,-i\}$. We set
$\te E=\tau\bar E\in\mor(I,\wb\tp\wb)$,
$\te E'=\ov{E'}\tau\in\mor(\wb\tp\wb,I)$, where $e_i\ot e_j$, $e^j\ot e^i$,
$i,j=1,2$, are treated as reals. Using \r{4.1},\r{4.2} and
\[ (E'\ot\jb)(\jb\ot E)=\jb,\qquad (\jb\ot E')(E\ot\jb)=\jb \]
(see (3) of \cite{WZ}; matrices of $E'$ and $E$ are inverse one to another),
one obtains (using e.g. diagram notation)
\be (\jb\ot X^{-1})(X^{-1}\ot\jb)(\jb\ot E)=E\ot\jb,\e{4.3}\ee
\be \beta^{-2}(\jb\ot X)(X\ot\jb)(\jb\ot\te E)=\te E\ot\jb,\e{4.4}\ee
\be \beta^2(X^{-1}\ot\jb)(\jb\ot X^{-1})(\te E\ot\jb)=\jb\ot\te E,\e{4.5}\ee
\be (\jb\ot E')(X\ot\jb)(\jb\ot X)=E'\ot\jb,\e{4.6}\ee
\be (E'\ot\jb)(\jb\ot X^{-1})(X^{-1}\ot\jb)=\jb\ot E',\e{4.7}\ee
\be \beta^{-2}(\te E'\ot\jb)(\jb\ot X)(X\ot\jb)=\jb\ot\te E',\e{4.8}\ee
\be \beta^2(\jb\ot\te E')(X^{-1}\ot\jb)(\jb\ot X^{-1})=
\te E'\ot\jb.\e{4.9}\ee

\bp\e{p7.0} (cf Theorem 6.3 of \cite{QDLG}, Remark 2 on page 229 of
\cite{WZ})\\
Let $q\in\cb\setminus\{0, \mbox{ roots of unity }\}$ (we treat $q=\pm1$ as
{\bf not} a root of unity). Then
\btr
\item[1)] there exist representations $w^s$ ($s\in\nb/2$) of $H$ such that
$w^0=I$, $w^{1/2}=w$, $\dim\ w^s=2s+1$ and
\[ w^s\tp w^{s'}\simeq w^{|s-s'|}\op w^{|s-s'|+1}\op\k\op w^{s+s'}\qquad
(s,s'\in\nb/2) \]
\item[2)] $w^s\tp\ov{w^{s'}}$ ($s,s'\in\nb/2$) are all unequivalent
irreducible representations of $H$
\item[3)] $w^s\tp\ov{w^{s'}}\simeq\ov{w^{s'}}\tp w^s$ ($s,s'\in\nb/2$)
\item[4)] each representation of $H$ is completely reducible\etr\ep

\bd Let $\api_{hol}$ be the subalgebra of $\api$ generated by matrix
elements of $w$. Then $Poly(H_{hol})=(\api_{hol},\de_{|_{\api_{hol}}})$ is
a Hopf subalgebra of $Poly(H)=(\api,\de)$. According to Proposition 4.1.1 of
\cite{WZ}, $\api_{hol}$ is the universal algebra generated by matrix
elements of $w$ satisfying the reletions \r{1.3} and \r{1.4}. Due to Theorem
4.2 of \cite{NQD} and the facts given in cases I,III of Introduction
to \cite{NQD} (cf (1.9), (1.30) and Theorem 1.15 of \cite{KP}), 1) holds and
matrix elements of $w^s$ ($s\in\nb/2$) form a linear basis of $\api_{hol}$.
Using Proposition 4.1.2--3 of \cite{WZ}, matrix elements of $w^s\tp\ov{w^{s'}}$
($s,s'\in\nb/2$) form a linear basis of $\api$. Now Proposition 4.1 of
\cite{NQD} (see also Proposition A.2 of \cite{P}) gives 2) and 4). The
condition iii. of Section 1 implies $(Tr\ w)(Tr\ \wb)=(Tr\ \wb)(Tr\ w)$.
That and 1) give that $Tr\ v$ ($v\in\Irr\ H$) commute among themselves. In
virtue of Proposition B.4 of \cite{KP}
(cf also Proposition 5.11 of \cite{W1}), one obtains 3).\ed

\noindent {\it Proof of Theorem \re{t1.2}.\/} We have Hopf ${}^*$-algebra
$\bpi$,
its Hopf ${}^*$-subalgebra $\api$ and two--dimensional representation $w$ of
$\api$ which satisfy the conditions i.--iv.,\r{1.2} and 1.--6. of Section 1.
We shall use the results of Section 1 of \cite{INH} with $\la$
replaced by $\lpi=w\tp\wb$, $\lpi_{AB,CD}=w_{AC}{w_{BD}}^*$.
Hence we deal with $p_{AB}=V_{AB,i}p_i$ instead of $p_i$.
In virtue of
(S1.3), it suffices to check (S1.5) for the generators: $a=w_{AB}$ or
$a={w_{AB}}^*$. Inserting such $a$ into (S1.5), we get
\[ G_{\lpi}\in\mor(w\tp w\tp\wb,w\tp\wb\tp w),\qquad
\te G_{\lpi}\in\mor(\wb\tp w\tp\wb,w\tp\wb\tp\wb), \mbox{ where } \]
\be (G_{\lpi})_{ABC,DEF}=f_{AB,EF}(w_{CD}),\qquad
(\te G_{\lpi})_{ABC,DEF}=f_{AB,EF}({w_{CD}}^*). \e{4.9a}\ee
Thus $G_{\lpi}=(\jb\ot X)A$, $\te G_{\lpi}=B(X^{-1}\ot\jb)$, where $A$ is an
intertwiner of $w\tp w\tp\wb\simeq w^1\tp\wb\op\wb$, $B$ is an intertwiner of
$w\tp\wb\tp\wb\simeq w\tp\ov{w^1}\op w$. But $w^1\tp\wb$, $\wb$,
$w\tp\ov{w^1}$, $w$ are irreducible (we use Propositions 4.1 and
4.2 of \cite{WZ}), hence
\[ \mor(w\tp w\tp\wb,w\tp w\tp\wb)=\cb EE'\ot\jb\op\cb\jb^{\ot 3}, \]
\[ \mor(w\tp\wb\tp\wb,w\tp\wb\tp\wb)=\cb\jb\ot\te E\te E'\op\cb\jb^{\ot 3}.
\]
Therefore $A=L\ot\jb$, $B=\jb\ot\te L$, where
\be L=a\jb^{\ot 2}+bEE',\quad \te L=\te a\jb^{\ot 2}+\te b\te E\te E',\quad
a,\te a,b,\te b\in\cb.\e{4.9a'}\ee

According to (S1.3), $f:\api\lra M_4(\cb)$ should be a unital homomorphism.
It means that $f$ should preserve the relations \r{1.3},
$\r{1.3}^*$, \r{1.4},
$\r{1.4}^*$, \r{1.5} ($\r{1.3}^*$ denotes the relation conjugated to \r{1.3}
etc.), i.e.
\be (G_{\lpi}\ot\jb)(\jb\ot G_{\lpi})(E\ot\jb^{\ot 2})=\jb^{\ot 2}\ot E,
\e{4.10}\ee
\be (\te G_{\lpi}\ot\jb)(\jb\ot\te G_{\lpi})(\te E\ot\jb^{\ot 2})=
\jb^{\ot 2}\ot \te E, \e{4.11}\ee
\be (\jb^{\ot 2}\ot E')(G_{\lpi}\ot\jb)(\jb\ot G_{\lpi})=E'\ot\jb^{\ot 2},
\e{4.12}\ee
\be (\jb^{\ot 2}\ot\te E')(\te G_{\lpi}\ot\jb)(\jb\ot\te G_{\lpi})=
\te E'\ot\jb^{\ot 2}, \e{4.13}\ee
\be (\te G_{\lpi}\ot\jb)(\jb\ot G_{\lpi})(X \ot\jb^{\ot 2})=
(\jb^{\ot 2}\ot X)(G_{\lpi}\ot\jb)(\jb\ot\te G_{\lpi}). \e{4.14}\ee
Using \r{4.3}, \r{4.5}, \r{4.6} and \r{4.8}, the equations \r{4.10}--\r{4.13}
are equivalent to
\be (L\ot\jb)(\jb\ot L)(E\ot\jb)=\jb\ot E,\e{4.15}\ee
\be \beta^{-2} (\te L\ot\jb)(\jb\ot\te L)(\te E\ot\jb)=
\jb\ot\te E,\e{4.16}\ee
\be (\jb\ot E')(L\ot\jb)(\jb\ot L)=E'\ot\jb,\e{4.17}\ee
\be\beta^{-2}(\jb\ot\te E')(\te L\ot\jb)(\jb\ot\te L)=
\te E'\ot\jb.\e{4.18}\ee
Using \r{4.9a'}, computing $a,b,\te a,\te b$, and
inserting them into \r{4.9a'},
one gets that the solutions of \r{4.15}--\r{4.18} are
\be L=L_i,\qquad\te L=\beta\tau\ov{L_j^{-1}}\tau,\qquad i,j=1,2,3,4,\e{4.19}
\ee
where
\be L_i=q_i(\jb^{\ot 2}+q_i^{-2}EE'),\e{4.20}\ee
$q_{1,2}=\pm q^{\frac12}$, $q_{3,4}=\pm q^{-\frac12}$.
Using these relations, \r{4.1}, \r{4.4}, \r{4.6} and \r{4.8},
we get that \r{4.14} is
satisfied. Therefore, the solutions of (S1.3),(S1.5) are given by
\r{4.9a},
where
\be G_{\lpi}=(\jb\ot X)(L\ot\jb),\qquad \te G_{\lpi}
=(\jb\ot\te L)(X^{-1}\ot\jb),\e{4.22}\ee
$L,\te L$ are given by \r{4.19}--\r{4.20} (in general 16 solutions).
Moreover,
\[ (R_{\lpi})_{ABCD,EFGH}=f_{AB,GH}(\lpi_{CD,EF})= \]
\[ f_{AB,MN}(w_{CE})f_{MN,GH}({w_{DF}}^*)=
(G_{\lpi})_{ABC,EMN}(\te G_{\lpi})_{MND,FGH}, \]
\be R_{\lpi}=(G_{\lpi}\ot\jb)(\jb\ot\te G_{\lpi})=
(\jb\ot X\ot\jb)(L\ot\te L)(\jb\ot X^{-1}\ot\jb).\e{4.23}\ee
We know that $\dot p_{AB}\ot\dot p_{CD}$ form a basis of $(\gd_2)_{inv}$
which transforms under $\de_{2L}$ according to $\lpi\tp\lpi$. It is easy to
check that the decomposition into irreducible unequivalent components
\[ \lpi\tp\lpi\simeq w\tp w\tp\wb\tp\wb\simeq
w^1\tp\ov{w^1}\op w^1\op\ov{w^1}\op I \]
corresponds to
\be (\gd_2)_{inv}=W_{1\bar 1}\op W_1\op W_{\bar 1}\op W_0 \e{4.23'}\ee
where
\[ W_{1\bar 1}=\spa\{(\phi\ot\psi)(\jb\ot X^{-1}\ot\jb)(\dot p\ot\dot p):
\ \phi,\psi\in(\cb^2\ot\cb^2)',\quad \phi E=0,\quad \psi\te E=0\}, \]
\[ W_{1}=\{(\phi\ot\te E')(\jb\ot X^{-1}\ot\jb)(\dot p\ot\dot p):
\ \phi\in(\cb^2\ot\cb^2)',\quad \phi E=0\}, \]
\[ W_{\bar 1}=\{(E'\ot\psi)(\jb\ot X^{-1}\ot\jb)(\dot p\ot\dot p):
\ \psi\in(\cb^2\ot\cb^2)',\quad \psi\te E=0\}, \]
\[ W_{0}=\cb(E'\ot\te E')(\jb\ot X^{-1}\ot\jb)(\dot p\ot\dot p) \]
(indices as in matrix multiplication rule have been omitted).
But $R^T_{\lpi}$ is the matrix of $\rho$ in the basis
$\dot p_{AB}\ot\dot p_{CD}$ (see remark after (S1.13)). Using \r{4.23},
we get that \r{4.23'} corresponds to
\[ \rho=\beta q_i\ov{q_j}^{-1}
\op -\beta q_i\ov{q_j}^3
\op -\beta q_i^{-3}\ov{q_j}^{-1}
\op\beta q_i^{-3}\ov{q_j}^3. \]

Comparing the condition 6. with Proposition {S1.6}, we get $\dim\ K=6$.
Therefore $K_{inv}=W_1\op W_{\bar 1}$. But Proposition {S1.4} implies
$K\subset\ker(\rho+\id)$, hence
$\beta q_i\ov{q_j}^3=\beta q_i^{-3}\ov{q_j}^{-1}=1$.
Remembering that $\beta\in\{1,-1,i,-i\}$, $q\neq\pm i$, we get
$q=\pm1$. Thus we can (and will) omit
$L_3,L_4$.
We obtain $\beta=q$, $i=j$ or $\beta=-q$, $i\neq j$ ($q\in\{1,-1\}$,
$i,j\in\{1,2\}$). In all these cases
$\rho=1\op-1\op-1\op1$,
hence
$K=\ker(\rho+\id)$. Moreover,
$\rho^2=\id$, $R^2=\jb^{\ot 4}$.
In virtue of Proposition \re{p7.0} the conditions a)--c) of Section 2
of \cite{INH} are
satisfied and we can use the results of Sections 1--4 of \cite{INH}.
In particular,
Corollary {S4.2} implies the first statement of the Theorem.

Let us pass from $\lpi$ to $\la=V^{-1}\lpi V$ (see \r{1.2}). Since $\bar
V=\tau V$, $\bar\la=\la$. We replace $p_{AB}$, $A,B=1,2$, corresponding to
$\lpi$ by $p_i=V^{-1}_{i,AB}p_{AB}$, $f_{AB,CD}$ by
$f_{ij}=V^{-1}_{i,AB}f_{AB,CD}V_{CD,j}$ (cf (S1.2)),
$R_{\lpi}$, $G_{\lpi}$ and $\te G_{\lpi}$ by
$R=(V^{-1}\ot V^{-1})R_{\lpi}(V\ot V)$,
$G=(V^{-1}\ot\jb)G_{\lpi}(\jb\ot V)$,
$\te G=(V^{-1}\ot\jb)\te G_{\lpi}(\jb\ot V)$.
Then \r{4.9a} gives
\be f_{ij}(w_{CD})=G_{iC,Dj},\quad f_{ij}(w_{CD}^*)=\te G_{iC,Dj}.\e{4.23''}\ee

Now we pass to a new $p_i$ such that $p_i^*=p_i$ without change of
$\tb^N$, $\gd_2$, $\xi$, $\rho$, $K$, $f_{ij}$, $R$, $G$ and $\te G$
(see Proposition {S4.5}.1 and (S1.2)). We redefine $p_{AB}$
accordingly.
In virtue of
Proposition {S4.5}.2, (S4.10) holds. Setting $a={w_{EF}}^*$ and passing
back
to $\lpi$ one has
$f_{BA,DC}(w^{-1}_{EF})=\ov{f_{AB,CD}({w_{EF}}^*)}$. It means
\be (G_{\lpi}^{-1})_{EBA,DCF}=\ov{(\te G_{\lpi})_{ABE,FCD}}, \e{4.23a}\ee
i.e. $(L_i^{-1}\ot\jb)(\jb\ot X^{-1})=(L_j^{-1}\ot\jb)(\jb\ot X^{-1})$ (we
used \r{4.22}, \r{4.19}, \r{4.2}). Thus $L_i=L_j$, $i=j$. Consequently,
$\beta=q=\pm1$ and $i=j=1,2$. Conversely, this condition gives (S4.10) for
$a={w_{EF}}^*$ and (using $S\circ *=*\circ S^{-1}$) $a=w^{-1}_{EF}$, hence
for all $a\in\api$.
The list of $X$ such that $\beta=q=\pm1$ is provided in the formulation of
Theorem \re{t1.2} (they contain factor $\al$ which is computed in such a way
that \r{4.1} is satisfied,
we also restricted the range of parameters according to remarks
on page 220 of \cite{WZ}). For $E,E',X$ as in Theorem \re{t1.2} and
$f_{ij}$ computed above
(S1.3), (S1.5) and (S4.10) (for $\la$) are satisfied.

According to Proposition \re{p7.0}, the only 2-dimensional
irreducible representations of $H$ are $UwU^{-1}$,
$U\wb U^{-1}$, $U\in GL(2,\cb)$. Thus if $\phi:\api_1\lra\api_2$ is an
isomorphism of Hopf ${}^*$-algebras $\api_1$, $\api_2$ included in our list,
then
\[ (1)\ \phi(w)=UwU^{-1}\ \mbox{ or }\ (2)\ \phi(w)=U\wb U^{-1}. \]
Let us consider the case (1).
We denote $E,E',X$ for $A_i$, $i=1,2$, by $E_i,{E_i}\pri,X_i$. Applying
$\phi$ to \r{1.3}--\r{1.5} for $\api_1$, one gets
\be (U^{-1}\ot U^{-1})E_1=k^{-1}E_2,\e{4.24}\ee
\be {E_1}\pri(U\ot U)=k'{E_2}\pri,\e{4.25}\ee
\be (\bar U^{-1}\ot U^{-1})X_1(U\ot\bar U)=lX_2\e{4.26}\ee
for some $k,k',l\in\cbg$. Considering \r{4.24}--\r{4.25}, one gets $E_1=E_2=E$
and
\[ U\in GL(2,\cb), k=k'=\det\ U \mbox{ for } E=e_1\ot e_2-e_2\ot e_1, \]
\[ U\in\{ \left(\ba{cc} m & x\\ 0  & m \ea\right): m\in\cbg,\quad x\in\cb\},
\quad k=k'=m^2 \]
for  $E=e_1\ot e_2-e_2\ot e_1+e_1\ot e_1$,
\[ U\in\{ \left(\ba{cc} x &0 \\ 0 & y \ea\right),
          \left(\ba{cc} 0 &x \\ y & 0 \ea\right): x,y\in\cbg\},\quad k=k'=xy \]
for $E=e_1\ot e_2+e_2\ot e_1$.

Inserting such $U$ in \r{4.26}, one gets (for $E=e_1\ot e_2-e_2\ot e_1$ see
Section 5.1 of \cite{WZ}) $X_1=X_2=X$, $l=1$, so
\be (\bar U^{-1}\ot U^{-1})X(U\ot\bar U)=X \e{4.28}\ee
and in particular cases:
\btr
\item[1)] $t=1$: $U\in GL(2,\cb)$\\
          $0<t<1$: $U\in\{ \left(\ba{cc} x & 0\\ 0 & y \ea\right),
                            \left(\ba{cc} 0 & x\\ y & 0 \ea\right):
x,y\in\cbg\}$
\item[2)] $U=m\left(\ba{cc} e^{i\phi} & x\\ 0 & e^{-i\phi} \ea\right)$,
          $m\in\cbg$, $x\in\cb$, $\phi\in\rb$
\item[3)] $U=m\left(\ba{cc} 1 & x\\ 0 & 1 \ea\right)$, $m\in\cbg$, $x\in\cb$
\item[4)] $U=m\left(\ba{cc} 1 & ix\\0  &1  \ea\right)$, $m\in\cbg$, $x\in\rb$
\item[5)] $U\in\{ \left(\ba{cc} x & 0\\ 0 & y \ea\right),
                  \left(\ba{cc} 0 & x\\ y & 0 \ea\right): x,y\in\cbg\}$
\item[6)] $U\in\left\{ m\left(\ba{cc} e^{i\phi} & 0\\
                                        0 & e^{-i\phi} \ea\right),
           \quad m\in\cbg,\quad \phi\in\rb\right\}$
\item[7)] $U\in\{ m\left(\ba{cc} 1 & 0\\ 0 & 1 \ea\right),\quad
                  m\left(\ba{cc} 1 & 0 \\ 0 & -1 \ea\right),\quad
                  m\left(\ba{cc} 0 & 1 \\ 1 & 0 \ea\right),\quad
                  m\left(\ba{cc} 0 & 1 \\ -1 & 0 \ea\right):\ m\in\cbg\}$.
\etr

Next, let us consider the case (2). Then
\be (\ov{U^{-1}}\ot\ov{U^{-1}})\te E_1=\bar k^{-1} E_2,\e{4.29}\ee
\be \te{E_1}\pri(\bar U\ot\bar U)=\bar k'{E_2}\pri,\e{4.30}\ee
\be (\bar U^{-1}\ot U^{-1})X_1(U\ot\bar U)=l^{-1}{X_2}^{-1}\e{4.31}\ee
for some $k,k',l\in\cbg$. Considering \r{4.29}--\r{4.30}, one gets
$E_1=E_2=E$,
\[ U\in\ GL(2,\cb),\ k=k'=-\det\ U \mbox{ for } E=e_1\ot e_2-e_2\ot e_1, \]
\[ U\in\{ \left(\ba{cc} m & x \\ 0 & -m \ea\right):\ m\in\cbg,
\quad x\in\cb\},\
k=k'=m^2, \]
for $E=e_1\ot e_2-e_2\ot e_1+e_1\ot e_1$,
\[ U\in\{ \left(\ba{cc} x & 0 \\ 0 & y \ea\right),\quad
          \left(\ba{cc} 0 & x \\ y & 0 \ea\right):\ x,y\in\cbg\}, k=k'=xy, \]
for $E=e_1\ot e_2+e_2\ot e_1$.

Inserting such $U$ in \r{4.31}, it is possible only for $X_1=X_2=X$ in the
following cases:
\btr
\item[1)] $t=1$: $U\in GL(2,\cb)$, $l=1$
\item[3)] $r=0$: $U=m\left(\ba{cc} 1 & x \\ 0 & 1 \ea\right)$, $m\in\cbg$,
$x\in\cb$,
$l=1$
\item[4)] $U=m\left(\ba{cc} 1 & \frac12 \\ 0 & -1 \ea\right)\cdot
              \left(\ba{cc} 1 & ix \\ 0 & 1 \ea\right)$, $m\in\cbg$,
$x\in\rb$, $l=1$
\item[5)] $t=1$: $U\in\{ \left(\ba{cc} x & 0 \\ 0 & y \ea\right),\quad
                         \left(\ba{cc} 0 & x \\ y & 0 \ea\right):
\ x,y\in\cbg\}$, $l=-1$
\etr
($l$ is computed for normalization of $X$ as in Theorem \re{t1.2}, which
includes $\al$).

In particular, all considered $(\api,\de)$ are nonisomorphic.\hfill$\Box$.\\

\br Using \r{4.15}--\r{4.19} for $i=j=1,2$, $\beta=q=\pm1$, one also gets
\be (\jb\ot L)(L\ot\jb)(\jb\ot E)=E\ot\jb,\e{4.31a}\ee
\be (\jb\ot\te L)(\te L\ot\jb)(\jb\ot\te E)=\te E\ot\jb,\e{4.31b}\ee
\be (E'\ot\jb)(\jb\ot L)(L\ot\jb)=\jb\ot E',\e{4.31c}\ee
\be (\te E'\ot\jb)(\jb\ot\te L)(\te L\ot\jb)=\jb\ot\te E'.\e{4.31d}
\ee\er

Let us repeat that $K=\ker(\rho+\id)$, $\rho^2=\id$, the conditions a)--c)
of Section~2 of \cite{INH} are satisfied and we can use the results of
Sections 1--4 of \cite{INH}.
We notice that $L,\te L,G,\te G$
and $R$ given before Theorem \re{t1.3} and in the
proof of Theorem \re{t1.2} coincide ($i=j=1$ corresponds to $\es=1$, while
$i=j=2$ to $\es=-1$). They correspond to $\la$ as in \r{1.2}, $\bar\la=\la$.

{\em Proof of Theorem \re{t1.3}.} Using Theorem \re{t1.2} and Corollary
{S3.8}.a, $\bpi$ is the universal
${}^*$-algebra with $I$ generated by $w_{AB}$ and
$p_i$ satisfying \r{1.3}--\r{1.8}. Next, (S2.6) coincides with a),
(S4.10)--(S4.11) imply b), \r{4.23''} gives the first formula of c).
The next two formulae in c)
can
be treated as definitions of $H_{EFCD}$ and $T_{EFCD}$. Since $w,\wb$ and
$\ppi$ are representations, formulae concerning Hopf structure follow.
Uniqueness of $f$, $\eta$, $T$ and $*$-Hopf
structure is obvious.

Let $\bpi,\hat\bpi$ describe two \qpg s and $\api,\de,\la,p,f,\eta,T$,
$\hat\api,\hat\de,\hat\la,\hat p,\hat f,\hat\eta,\hat T$ be the corresponding
objects as in Theorems \re{t1.2} and \re{t1.3}. Assume that
$\phi:\bpi\lra\hat\bpi$ is an isomorphism of Hopf ${}^*$-algebras. According
to Proposition {S4.4}, one has $\phi(\api)=\hat\api$ and we put
$\phi_{\api}=\phi_{|_{\api}}:\api\lra\hat\api$. Due to the proof of Theorem
\re{t1.3}, one has
\[ (1)\ \phi(w)=U\hat w U^{-1}\mbox{ or } (2)\ \phi(w)=U\ov{\hat w}U^{-1}. \]
Using \r{1.2}, one gets $\phi(\la)=M\hat\la M^{-1}$ where
\be\left.\ba{c} M=V^{-1}(U\ot\bar U)V \mbox{ in the case (1), }\\\\
  M=q^{1/2}V^{-1}(U\ot\bar U)XV \mbox{ in the case
(2). }\ea\right\}\e{til}\ee
Using \r{1.2a} and \r{4.2}, one gets $\bar M=M$ ($\ov{q^{1/2}}=\beta q^{1/2}$
since $\beta=q=\pm1$).

In virtue of (S4.2), one has
\[ \hat f_{ij}(\hat w_{CD})
=U^{-1}_{CA}(M^{-1})_{il}f_{lm}(w_{AB})M_{mj}U_{BD}
\mbox{ in the case (1), } \]
\[ \hat f_{ij}(\hat w_{CD})=
\bar U^{-1}_{CA}(M^{-1})_{il}f_{lm}({w_{AB}}^*)M_{mj}\bar U_{BD}
\mbox{ in the case (2). } \]
Using \r{4.24}--\r{4.26} or
\r{4.29}--\r{4.31}, we get $\hat L=L$ in all cases. Thus
there are no isomorphisms between \qpg s with different $\es$.\hfill$\Box$.\\

Using the computer MATHEMATICA program, we made several
computations performed in

{\em Proof of Theorem \re{t1.4}.} Let
$\bpi,\api,\de,\la,p,f,\eta,T$ describe a \qpg. According to Propositions
{S4.4} and {S4.5}.3, it is always possible to replace $\eta$ by
$\hat\eta=\eta+fh-\eps h$
where $h_i\in\rb$. We put
$\api=\hat\api$, $w=\hat w$, $c=1$,
$M=\jb_4$, $\phi_{\api}=\id$, and
$f$ doesn't change. Thus we substitute $H_{EFCD}$ by
\[ \hat H_{EFCD}=V_{EF,i}\hat\eta_i(\hat w_{CD})=
H_{EFCD}+f_{EF,AB}(w_{CD})h_{AB}-\del_{CD}h_{EF}, \]
where $f_{EF,AB}(w_{CD})$ are given by \r{4.9a} and \r{4.22},
$h_{EF}=V_{EF,i}h_i$ (i.e. $h_{11},h_{22}\in\rb$, $\ov{h_{12}}=
h_{21}\in\cb$).
In each equivalence class obtained by such substitutions we restrict
ourselves
to exactly one $H$ singled out by the following constraints:
\btr
\ite no constraints for 1), $\es=1$, $t=1$,
\ite $H_{1111}\in i\rb$, $H_{2222}\in i\rb$, $H_{1222}=0$ for 1), $\es=1$,
$t\neq1$,
\ite $H_{1112}=0$, $H_{2112}\in i\rb$ for 2), $\es=1$,
\ite $H_{1111}=0$, $H_{1112}\in i\rb$, $H_{1211}\in i\rb$ for 3), $\es=1$,
\ite $H_{2111}\in i\rb$, $H_{1122}\in\rb$, $H_{1112}\in i\rb$ for 4),
$\es=1$,
\ite $H_{1111}\in i\rb$, $H_{2111}=0$, $H_{2222}\in i\rb$ for 5), $\es=1$,
$t\neq1$,
\ite $H_{1111}\in i\rb$, $H_{2222}\in i\rb$ for 5), $\es=1$, $t=1$,
\ite $H_{1122}\in i\rb$, $H_{1112}=0$, $H_{2211}\in i\rb$ for 6), $\es=1$,
\ite $H_{1122}\in i\rb$, $H_{2222}\in i\rb$, $H_{1222}=0$ for 7), $\es=1$,
\ite $H_{1111}\in i\rb$, $H_{1222}=0$, $H_{2222}\in i\rb$ for 1), $\es=-1$,
\ite $H_{1122}\in i\rb$, $H_{1211}=0$, $H_{2211}\in i\rb$ for 2), $\es=-1$,
\ite $H_{2111}=0$, $H_{1111}\in i\rb$, $H_{2211}\in i\rb$ for 3), $\es=-1$,
\ite $H_{2211}\in i\rb$, $H_{1222}=0$, $H_{1111}\in i\rb$ for 4), $\es=-1$,
\ite $H_{1222}=0$, $H_{1111}\in i\rb$, $H_{2222}\in i\rb$ for 5), $\es=-1$,
$t\neq1$,
\ite $H_{1222}=0$ for 5), $\es=-1$, $t=1$,
\ite $H_{1211}=0$, $H_{2112}\in i\rb$ for 6), $\es=-1$,
\ite $H_{1122}\in i\rb$, $H_{1222}=0$, $H_{2222}\in i\rb$ for 7), $\es=-1$.
\etr
We also may and will assume (S3.50).

In virtue of the theory presented in
Sections 1--4 of \cite{INH}
(see e.g. Theorem {S3.1} and Proposition {S4.5})
$H_{EFCD}$ and $T_{EFCD}$ give a \qpg\ if and only if (S1.5), (S2.6),
(S2.14), (S3.1), (S3.2), (S4.10), (S4.11) and (S4.12) are satisfied
(cf the proof of Theorem \re{t1.2}). We shall investigate subsequent
conditions and dealing with next ones we assume that previously investigated
conditions are satisfied.
We
already know that $f$ is a unital homomorphism satisfying
\r{4.23''}, (S1.5) and
(S4.10).
Thus (S2.6) means that applying $\eta_i$ to the relations \r{1.3},
$\r{1.3}^*$, \r{1.4}, $\r{1.4}^*$ and \r{1.5} ($*$ means that we conjugate
the
relation) and using (S2.5), one gets relations on
$H^w_{iA,B}=\eta_i(w_{AB})$ and $H^{\wb}_{iA,B}=\eta_i({w_{AB}}^*)$, which
should be satisfied. They read as follows:
\be \{(G\ot\jb)(\jb\ot H^w)+(H^w\ot\jb)\}E=0,\e{4.32} \ee
\be \{(\te G\ot\jb)(\jb\ot H^{\wb})+(H^{\wb}\ot\jb)\}\te E=0,\e{4.33} \ee
\be (\jb\ot E')\{(H^w\ot\jb)+(G\ot\jb)(\jb\ot H^w)\}=0,\e{4.34}\ee
\be (\jb\ot\te E')\{(H^{\wb}\ot\jb)+(\te G\ot\jb)(\jb\ot H^{\wb})\}
=0,\e{4.35}
\ee
\be\left.\ba{c}
(\jb^{\ot 2}\ot X)\{(H^w\ot\jb)+(G\ot\jb)(\jb\ot H^{\wb})\}=\\\\
\{(H^{\wb}\ot\jb)+(\te G\ot\jb)(\jb\ot H^w)\}X.\ea\right\} \e{4.36}\ee
Setting $a={w_{EF}}^*$ in (S4.11), one gets
\be (H^{\wb})_{iE,F}=\ov{\eta_i(w^{-1}_{EF})}=
-\ov{f_{ij}(w^{-1}_{EL})}\cdot\ov{\eta_j(w_{LF})}=
-\ov{G^{-1}}_{Ei,jL}\cdot\ov{H^w_{jL,F}}\e{4.37}\ee
(we used (S2.5), (S1.4) and \r{4.23''}). Conversely, \r{4.37} gives
(S4.11)
for $a={w_{EF}}^*$ and (using $S\circ *\circ S\circ *=\id$)
$a=w^{-1}_{EF}$, hence for all $a\in\api$ (it suffices to check the
conditions
(S4.10)--(S4.11) on generators of $\api$ as algebra: they are equivalent to
Theorem \re{t1.3}.b for $a^*$).

Using the 16 relations \r{4.1}, \r{4.3}--\r{4.9}, \r{4.15}--\r{4.18} and
\r{4.31a}--\r{4.31d}, one gets that \r{4.32} is equivalent to \r{4.34},
\r{4.33} is equivalent to \r{4.35}. Moreover, \r{4.32} is equivalent to
\r{4.33} (one conjugates \r{4.33} and uses \r{4.23a}, \r{4.37}). Thus
\r{4.33}--\r{4.35} are superfluous. The remaining equations: \r{4.36} (with
inserted \r{4.37}) and \r{4.32} give a set of $\rb$-linear equations on
$H_{EFCD}=V_{EF,i}H^w_{iC,D}$.

Next, (S3.50) gives a set of linear equations on
$T_{EFCD}=V_{EF,i}V_{CD,j}T_{ij}$.
By virtue of (S3.50) and (S4.14), one obtains ($\te T$ was defined
after (S4.12))
$R\te T=-\te T$, $RD=-D$, where $D=\te T-T$. Therefore $D$
corresponds to a subrepresentation of $\la\tp\la$ equivalent to
$w^1\op\ov{w^1}$. But (S4.12) means that $D$ is an invariant vector of
$\la\tp\la$, hence $D=0$, $\te T=T$ (conversely, this implies (S4.12)). This
gives a set of $\rb$-linear conditions on
$T_{EFCD}$.

According to Proposition {S3.13} and Corollary {S4.9}, we may replace
(S2.14)
by (S3.55) for $b=w_{AB}$. But this is equivalent to
$M\in\mor(w,\la\tp\la\tp w)$, where $M_{ijC,B}=\tau^{ij}(w_{CB})$. Using
$[(R+\jb_4^{\ot 2})\ot\jb]M=0$ (see (S3.54)), one gets
$M=[(V^{-1}\ot V^{-1})(\jb\ot X\ot\jb)\ot\jb]N$, where
$N\in\mor(w,w\tp w\tp\wb\tp\wb\tp w)$, $(L_1\ot\te L_1\ot\jb)N=-N$
($L\ot\te L$ doesn't depend on $\es$, one can put $\es=1$). Thus
$N_{ABCDF,G}=P_{ABF,G}\te E_{CD}$ with $P\in\mor(w,w\tp w\tp w)$,
$(L_1\ot\jb)P=q^{1/2}P$. It means $P=\lam\jb\ot E+\mu E\ot\jb$,
$(EE'\ot\jb)P=0$. Hence $\mu=\frac12 q\lam$,
$P=\lam(\jb\ot E+\frac12 qE\ot\jb)$. On the other hand,
\[ \tau^{ij}(w_{CB})=(R-\jb^{\ot 2})_{ij,kl}(\eta_l(w_{CA})\eta_k(w_{AB})- \]
\[ \eta_k(\la_{ls})\eta_s(w_{CB})+T_{kl}\del_{CB}-
f_{ln}(w_{CA})f_{km}(w_{AB})T_{mn}). \]
Using \r{1.2}, (S2.5), \r{4.37} and \r{4.23''}, one gets a set of equations
containing terms bilinear in $\Ren\ H_{ABCD}$, $\Imn\ H_{ABCD}$, terms linear
in
$\Ren\ T_{ABCD}$, $\Imn\ T_{ABCD}$ and terms linear in $\Ren\ \lam$,
$\Imn\ \lam$.

We shall prove that (S3.1) is equivalent to $\lam\in q^{1/2}\rb$.
One has (see
\r{1.2})
$\te F=(V^{-1}\ot V^{-1}\ot V^{-1})JV$ where
$\te F=((R-\jb_4^{\ot 2})\ot\jb)F$ and
\[ J_{QRTVAB,CD}=V_{QR,i}V_{TV,j}\tau^{ij}(w_{AC}{w_{BD}}^*). \]
Using Proposition {S3.13},
\[ \tau^{ij}(w_{AC}{w_{BD}}^*)=
\tau^{ij}(w_{AC})\del_{BD}+G_{jA,Es}G_{iE,Cm}\tau^{ms}({w_{BD}}^*). \]
But in virtue of Proposition {S4.8}
\[ \tau^{ms}({w_{BD}}^*)=\ov{\tau^{sm}(w^{-1}_{BD})}. \]
Using once again Proposition {S3.13} for $a=w_{LS}$, $b=w^{-1}_{SD}$,
multiplying both sides by $G^{-1}_{Bs,iP}G^{-1}_{Pm,jL}$ and conjugating both
sides, one gets
\[ \ov{\tau^{sm}(w^{-1}_{BD})}=
-\te G_{sB,Pi}\te G_{mP,Lj}\ov{\tau^{ij}(w_{LD})} \]
(see \r{4.23a}). Inserting all these data, after some calculations (using the
16 relations), one obtains
\[ J=\bar\lam q A+\lam B+\frac12 (\bar\lam+\lam q)C \]
\[ \mbox{ where }\quad A=(\jb\ot X\ot X\ot\jb)(E\ot X\ot\te E), \]
\[ B=\jb\ot (\jb\ot X^{-1}\ot\jb)(\te E\ot E)\ot\jb,\quad
C=(\jb\ot X\ot\jb)(E\ot\te E)\ot\jb\ot\jb. \]
We shall also use $D=\jb\ot\jb\ot(\jb\ot X\ot\jb)(E\ot\te E)$.
Using \r{4.23} and the 16
relations, one has
\[ (R_{\lpi}\ot\jb)A=-A-qC,\qquad (R_{\lpi}\ot\jb)B=-B-qC, \]
\[ (R_{\lpi}\ot\jb)C=C,\qquad (R_{\lpi}\ot\jb)D=D+qA+qB+C, \]
\[ (\jb\ot R_{\lpi})A=-A-qD,\qquad (\jb\ot R_{\lpi})B=-B-qD, \]
\[ (\jb\ot R_{\lpi})C=C+qA+qB+D,\qquad (\jb\ot R_{\lpi})D=D. \]
In particular, $(R_{\lpi}\ot\jb)J=-J$ (it also follows from (S3.54)). Thus
we
can compute
\[ -2(V\ot V\ot V)A_3FV^{-1}=(V\ot V\ot V)A_3\te FV^{-1}= \]
\[ [\jb\ot\jb\ot\jb-\jb\ot R_{\lpi}-R_{\lpi}\ot\jb+
(\jb\ot R_{\lpi})(R_{\lpi}\ot\jb)+ \]
\[ (R_{\lpi}\ot\jb)(\jb\ot R_{\lpi})-
(R_{\lpi}\ot\jb)(\jb\ot R_{\lpi})(R_{\lpi}\ot\jb)]J= \]
\[ 2[J-(\jb\ot R_{\lpi})J+(R_{\lpi}\ot\jb)(\jb\ot R_{\lpi})J]=
3(\bar\lam q-\lam)(A-B). \]
But $A\neq B$
($\im[(\jb\ot X^{-1}\ot X^{-1}\ot\jb)A]=\im E\ot\cb^2\ot\cb^2\ot \im\te E$
while $\im[(\jb\ot X^{-1}\ot X^{-1}\ot\jb)B]=\cb^2\ot W_0\ot\cb^2$,
where
$\dim\ \im E=\dim\ \im\te E=\dim\ W_0=1$), hence $A_3F=0$ if and only if
$\bar\lam=q\lam$, i.e. $\lam\in q^{1/2}\rb$.

We notice that
\[ \la\tp\la\tp\la\simeq w\tp w\tp w\tp\wb\tp\wb\tp\wb\simeq
(w\op w\op w^{3/2})\tp(\wb\op\wb\op\ov{w^{3/2}}), \]
hence
\be \mor(I,\la\tp\la\tp\la)=\{0\}.\e{4.37'}\ee
Therefore (S3.2) is equivalent to
$A_3(Z\ot\jb_4-\jb_4\ot Z)T=0$. This gives a set of equations, which are
$\rb$-linear in $\Ren\ H_{ABCD}\cdot \Ren\ T_{EFGH}$,
$\Imn\ H_{ABCD}\cdot \Ren\ T_{EFGH}$, $\Ren\ H_{ABCD}\cdot \Imn\ T_{EFGH}$ and
$\Imn\ H_{ABCD}\cdot \Imn\ T_{EFGH}$. Our strategy is as follows: we set
constraints for $H_{ABCD}$ as before, solve $\rb$-linear equations,
insert these data into $\rb$-bilinear equations
and finally use the condition for $\lam$ and the last set of equations.
In the cases 1), $t=1$, $\es=1$ and 5),
$t=1$, $\es=\pm1$, we haven't solved the $\rb$-bilinear equations
(but see Remark \re{r1.4}).
In other cases one gets the following solutions (with the parameters
being real numbers):
\btr
\ite \r{1.8a} with $T_{ab}=-T_{ba}\in i\rb$ for 1), $\es=-1$, $t=1$,
\ite \r{1.9} for 1), $\es=\pm1$, $0<t<1$,
\ite \r{1.10} or \r{1.11} for 2), $\es=1$,
\ite \r{1.10} with $a=b=0$ for 2), $\es=-1$,
\ite \r{1.12} for 4), $\es=1$,
\ite \r{1.13} for 5), $\es=\pm1$, $0<t<1$,
\ite \r{1.14} or \r{1.15} for 6), $\es=-1$,
\etr
in the remaining cases all $H_{EFCD}$ and $T_{EFCD}$ must equal $0$. Moreover,
$\lam=8b^2$ in the case 4), $\es=1$ and $\lam=0$ in other solved cases.

Let us remark that for fixed $w_{AB}$ and $p_i$: $\phi_i$, $\eta_i$ and
$H_{EFCD}$ are uniquely determined (cf (S1.6)). Moreover, $T_{EFCD}$
satisfying (S3.50) are also uniquely determined: if $T'$ would also satisfy
(S3.46) and (S3.50), then for $L=T-T'$ we would have
\[ 0=(R-\jb_4^{\ot 2})(L-(\la\tp\la)L)=(R-\jb_4^{\ot 2})L-
(\la\tp\la)(R-\jb_4^{\ot 2})L=-2(L-(\la\tp\la)L), \]
$L\in\mor(I,\la\tp\la)$, but $RL=-L$ gives that $L$ corresponds to the
subrepresentation $w^1\op\ov{w^1}$ of $\la\tp\la$, $L=0$, $T=T'$.

It remains to check which pairs $(H,T)$ as above give isomorphic objects.
In virtue of Propositions {S4.4} and {S4.5} and above remarks it would
mean that $(\hat H,\hat T)$ is obtained from $(H,T)$ via formulae
(S4.3)--(S4.4) with $c,h_i\in\rb$, $c\neq0$, $M$ as in \r{til}. After some
calculations one can choose one pair $(H,T)$ in each equivalence class (for
each considered case). The results are presented in the formulation of the
Theorem.\hfill$\Box$.\\

{\em Proof of Theorem \re{t1.5}.} In virtue of Corollary {S3.6} it
suffices to prove $\dim\ S_n=d_n$. Taking $\la=\lpi=w\tp\wb$, one has the
projection $S_n=\frac{1}{n!}\sum_{\pi\in\Pi_n}R_{\pi}$, where
$R_{\pi}=(R_{\lpi})_{i_1}\cdot\k\cdot(R_{\lpi})_{i_k}$ for a minimal
decomposition $\pi=t_{i_1}\cdot\k\cdot t_{i_k}$,
$R_{\lpi}=(\jb\ot X\ot\jb)(L\ot\te L)(\jb\ot X^{-1}\ot\jb)$. Putting
\[ K_X=(\jb^{\ot n-1}\ot X\ot\jb^{\ot n-1})
(\jb^{\ot n-2}\ot X\ot X\ot\jb^{\ot n-2})\cdot\k\cdot \]
\[ (\jb\ot X\ot\k\ot X\ot\jb)(X\ot\k\ot X) \]
and $K_{\tau}$ defined similarly with $X$ replaced by $\tau$, one can define
${S_n}\pri=K_XS_nK_X^{-1}$ and ${S_n}''=K_{\tau}^{-1}{S_n}\pri K_{\tau}$.
Therefore $\dim\ S_n=\tr\ S_n=\tr\ {S_n}\pri=\tr\ {S_n}''$.
One gets the formula for ${S_n}''$ as for $S_n$ but with
$R_{\lpi}$ replaced by
${R_{\lpi}}''=(\jb\ot\tau\ot\jb)(L\ot\te L)(\jb\ot\tau^{-1}\ot\jb)$ (we
use the
16 relations).

Moreover, $L\ot\te L=L_0\ot \tau L_0\tau$ where $L_0=\jb^{\ot 2}+q^{-1}EE'$,
$E=e_1\ot e_2-qe_2\ot e_1+t_0 e_1\ot e_1$,
$E'=-qe^1\ot e^2+e^2\ot e^1+t_0 e^2\ot e^2$, $q=\pm1$, $t_0=0,1$ (for $q=-1$
one has $t_0=0$). Replacing $e_1$ by $ce_1$, $c\neq0$, one has to replace
$e^1$ by $c^{-1}e^1$, $L_0$ by $L_0$ with $t_0$ replaced by $c\cdot t_0$.
Thus
(for $q=1$)
$\tr\ {S_n}''$ doesn't depend on $t_0\in\cb$ (for $t_0\neq0$ and also for
$t_0=0$ in limit). So we may put $t_0=0$. Then $L_0e_1\ot e_1=e_1\ot e_1$,
$L_0e_2\ot e_2=e_2\ot e_2$, $L_0e_1\ot e_2=qe_2\ot e_1$,
$L_0e_2\ot e_1=qe_1\ot e_2$. Setting $A_{\al\beta}=e_{\al}\ot e_{\beta}$,
one has
\[ {R_{\lpi}}''A_{\al\beta}\ot A_{\ga\del}=
q^{\al+\beta+\ga+\del}A_{\ga\del}\ot
A_{\al\beta}. \]
It is easy to show that ${S_n}''({R_{\lpi}}'')_k=({R_{\lpi}}'')_k{S_n}''=
{S_n}''$,
${S_n}''$ is a projection,
\[
{S_n}''(A_{11}^{\ot a}\ot A_{12}^{\ot b}
\ot A_{21}^{\ot c}\ot A_{22}^{\ot d}),
\qquad a+b+c+d=n, \]
form a basis of $\im\ {S_n}''$. We get
\[ \dim\ S_n=\tr\ {S_n}''=\dim\ \im\ {S_n}''= \]
\[  \#\{(a,b,c,d)\in\nb^{\ot 4}:
\ a+b+c+d=n\}=d_n. \]
\hfill$\Box$.\\

{\em Proof of Theorem \re{t1.7}.} We know that
$\la\tp\la\simeq I\op w^1\tp\ov{w^1}\op w^1\op\ov{w^1}$, where
$\ker(R+\jb_4^{\ot 2})$ corresponds to $w^1\op\ov{w^1}$.
Therefore (S5.2) holds.
Moreover, \r{4.37'} coincides with (S5.4).
Using Theorem {S5.6}, we get the first
statement. The second statement follows from Proposition {S5.3},
Proposition {S5.5} and
$\dim\ S_n=d_n$ (see the proof of Theorem \re{t1.5}).\hfill$\Box$.\\

{\em Proof of Theorem \re{t1.8}.} We know
that (S3.59) holds (see (S3.2) and \r{4.37'}) and $R\neq\pm\jb_4^{\ot 2}$
(see the proof of Theorem
\re{t1.2}).
Moreover,
$(\la\tp\la)m'=m'$ means that $m'$ is proportional to $m$.
According to the proof of Theorem
\re{t1.4}, $\te F=0$ if and only if $\lam=0$ (otherwise,
using $\bar\lam=q\lam$,
$A+B+qC=0$, acting $\jb\ot R_{\lpi}$, $C=D$,
$V_0\ot\cb^4=\im\ C=\im\ D=\cb^4\ot V_0$ where
$V_0=\im[(\jb\ot X\ot\jb)(E\ot\te E)]$, $\dim\ V_0=1$, contradiction), which
means $b=0$ in the case 4), $\es=1$ and no condition in other cases
listed in Theorem \re{t1.4}. Then we
use Proposition {S3.14}.\hfill$\Box$.\\

\br According to Corollary {S3.8}.b, $\bpi$ is the
universal unital algebra generated by $\api$ and $p_i$ ($i\in\spi$)
satisfying $I_{\bpi}=I_{\api}$,
(S3.48) and (S3.47) for $w$ and $\wb$
(cf Remark {S3.10}).\er

\begin{center} {\bf Acknowledgements} \end{center}

The first author is grateful to Prof. W. Arveson and other faculty members
for their kind hospitality in UC Berkeley.
The authors are thankful to Dr S. Zakrzewski for fruitful discussions.
\\\\


\begin{thebibliography}{xx}
\bibitem{ChD} Chaichian, M. and Demichev, A.P.,
Quantum Poincar\'e group,
{\it Phys. Lett.\/}
{\bf B304} (1993), 220--224. Cf also Schirrmacher, A., ``Varieties on quantized
spacetime'' in: ``Symmetry methods in physics'', A.N. Sissakin et al. (eds.),
vol. 2, 463--470, Dubna 1994.
\bibitem{D} Dobrev, V.K., Canonical $q$-deformations of noncompact Lie (super-)
algebras, {\it J.Phys.A: Math. Gen.\/} {\bf 26}(1993), 1317--1334.
\bibitem{KP} Kondratowicz, P. and Podle\'s, P., Irreducible representations
of quantum $SL_q(2)$ groups at roots of unity, hep-th 9405079.
\bibitem{L} Lukierski, J., Nowicki, A. and Ruegg, H., New quantum Poincar\'e
algebra and $\kappa$--deformed field theory, {\it Phys. Lett.\/}
{\bf B293} (1992), 344--352;
Zakrzewski, S., Quantum Poincar\'e group related to the $\kappa$-Poincar\'e
algebra, {\it J. Phys. A: Math. Gen.\/} {\bf 27} (1994), 2075--2082.
Cf also Lukierski,J., Nowicki, A., Ruegg, H. and Tolstoy, V.N.,
$q$-deformation of Poincar\'e algebra {\it Phys. Lett.\/} {\bf B264} (1991),
331--338.
\bibitem{M} Majid, S., Braided momentum in the $q$-Poincar\'e group,
{\it J. Math. Phys.\/} {\bf 34} (1993), 2045--2058.
\bibitem{O} Ogievetsky, O., Schmidke, W.B., Wess, J. and Zumino, B.,
$q$-Deformed Poincar\'e algebra, {\it Commun. Math. Phys.\/} {\bf 150}
(1992), 495--518.
\bibitem{P} Podle\'s, P., Complex quantum groups and their real
representations, {\it Publ. RIMS, Kyoto University\/} {\bf 28} (1992),
709--745.
\bibitem{QDLG} Podle\'s, P. and Woronowicz, S.L., Quantum
deformation of Lorentz group, {\it Commun. Math. Phys.\/} {\bf 130} (1990),
381--431.
\bibitem{PW} Podle\'s, P. and Woronowicz, S.L., Inhomogeneous quantum groups,
submitted to {\it Proceedings of First Caribbean School
of  Mathematics and Theoretical
Physics in Guadeloupe, 1993\/}.
\bibitem{INH} Podle\'s, P. and Woronowicz, S.L., On the
structure of inhomogeneous quantum groups,
hep-th 9412058, UC Berkeley preprint.
\bibitem{S} Schlieker, M., Weich, W. and Weixler, R., Inhomogeneous
quantum groups, {\it Z. Phys. C. -- Particles and Fields\/} {\bf 53} (1992),
79-82; Inhomogeneous quantum groups and their universal enveloping algebras,
{\it Lett. Math. Phys.\/} {\bf 27} (1993), 217--222.
\bibitem{W1} Woronowicz, S.L., Compact matrix pseudogroups, {\it Commun.
Math. Phys.\/} {\bf 111} (1987), 613--665.
\bibitem{NQD} Woronowicz, S.L., New quantum deformation of $SL(2,\cb)$.
Hopf algebra level, {\it Rep. Math. Phys.\/} {\bf 30(2)} (1991), 259--269.
\bibitem{WZ} Woronowicz, S.L. and Zakrzewski, S., Quantum deformations of the
Lorentz group.
The Hopf ${}^*$-algebra level, {\it Comp. Math.\/} {\bf 90} (1994),
211--243.
\bibitem{GQ} Zakrzewski, S., Geometric quantization of Poisson groups --
diagonal and soft deformations, {\it Contemp. Math.\/}, vol. 179 (1994),
271--285.
\bibitem{Z} Zakrzewski, S., Poisson Poincar\'e groups, submitted to
{\it Proceedings of Winter School of Theoretical Physics, Karpacz 1994},
hep-th 9412099;
Cf also Poisson homogeneous spaces, submitted to
{\it Proceedings of Winter School of Theoretical Physics, Karpacz 1994},
hep-th 9412101;
Poisson structures on the Poincar\'e group, in preparation.
\end{thebibliography}
\end{document}